\documentclass[
journal=nanolett, 
manuscript=article,
layout=twocolumn]{achemso}
\usepackage[version=3]{mhchem} 
\usepackage{color}
\usepackage[dvipsnames]{xcolor}
\usepackage{array}
\usepackage{ctable}

\author{Daryoush Shiri}
\author{Andreas Isacsson}

\affiliation[Chalmers]
{Department of Physics, Chalmers University of Technology, SE-412 96 G\"oteborg, Sweden}
\email{andreas.isacsson@chalmers.se}
\title{Dynamic in situ Control of Heat Rectification in Graphene Nano Ribbons using Electric Field-induced Strain }
\let\oldmaketitle\maketitle
\let\maketitle\relax

\begin{document}
\twocolumn[
  \begin{@twocolumnfalse}
    \oldmaketitle
    \begin{abstract}
An increasing number of papers propose routes to implement thermal counterparts of electronic rectification. These schemes are mainly based on combinations of crystal anharmonicity and broken mirror symmetry. With respect to graphene, proposals pivot around shape asymmetry induced by using hetero-structures of nano-patterned or defected sections of pristine graphene. Using Molecular Dynamics (MD) we show that it suffices to split a graphene nano-ribbon into two unequal strained sections using external force which leads to large asymmetry in the forward and reverse heat fluxes. We find that the corresponding rectification ratio is enhanced by up to 60~\%. 
Also, and more importantly, the polarity is controllable on-the-fly i.e. by changing the position where force is applied. Based upon our results we propose a thermo-electric device which obviates the complex nano-patterning and lithography required to pattern graphene every time a new rectification value or sign is sought for, opening a route to simpler fabrication and characterization of phononic phenomena in 2D materials. 

    \end{abstract}
  \end{@twocolumnfalse}
  ]

{\bf Introduction}. Graphene is currently a prime candidate material for micro-scale and nano-scale heat management applications~\cite{Cahill_rev2014} due to its exceptional thermal conductivity~\cite{Pop_MRS2012, Balandin_rev, Balandin2011}. By nanoscale patterning or shaping, the material properties can be tailored to suit particular demands. In the context of thermal transport, this pattering is done to locally modify the phonon dispersion relation~\cite{NuoYang_iop2016} and the corresponding research area, phononics, is sharply on the rise~\cite{Maldovan2013, RevModPhys_Li}. Phononics aims at engineering devices to control transport and detect phonons (heat carriers) in the same way electrons carry information in electronics. Just as the semiconductor diode paved the way for integrated electric circuitry, efforts to construct efficient acoustic and thermal diodes aim at a similar revolution in phononics~\cite{Maldovan2013}. 

In a thermal diode, or rectifier, the magnitude of the heat flux resulting from a temperature difference applied at the two terminals differ depending on which terminal is the hotter~\cite{RevModPhys_Li, Shih2015_maxRR}. An important figure of merit is the rectification ratio, $RR$, defined as the relative difference between these two heat fluxes (currents) or $RR= (J^{+}-J^{-})/J^{-}$ in which $J^{+}$ and $J^{-}$ are left to right (forward) and right to left (reverse) heat fluxes, respectively. However, in contrast to acoustic diode proposals, which mostly rely on single-mode phononic analogs of photonic crystals, the implementation of thermal diodes is not as straightforward. This is because heat is carried by a wide spectrum of phonon modes.

The necessary conditions for a device to display rectification are nonlinearity of the inter-atomic potential and broken mirror symmetry. If this leads to a temperature dependent mismatch of the vibrational density of states (vDOS) between the left and right regions, thermal rectification may ensue. Most proposals relating to carbon nanotubes~\cite{Wu_2007,Ni_2011} and graphene based thermal rectifiers have thus far focused on symmetry-breaking geometries which may be hard to realize experimentally~\cite{NuoYang_asym,Series_diode,Hu_2009,Yang_2009, Xu_2014,YanWang_nl14}.

Here we show that simply using two weakly connected graphene ribbon segments of unequal length can provide sufficient mismatch of vDOS for significant thermal rectification to emerge. We attribute this to the temperature dependency of the vDOS which emanates from the non-linearity of atomic potentials as well as the length difference. This leads to different vDOS overlaps for thermal bias in the forward and the backwards directions. This, in turn, results in a highly asymmetric flow of heat. Chief among our results is that the rectification ratio and its sign can be adjusted by changing the length of the short and long nano ribbon sections on-the-fly using electric field-induced strain. The value of $RR$ can reach up to 60~\% which is comparable or better than some of the proposals based on more elaborate nano-patterning of graphene. To put our results in perspective, a few graphene based rectifier proposals along with predicted values of $RR$ are listed in Table~\ref{table:RR}. 

As shown in Table~\ref{table:RR}, reported values of $RR$ vary among different works. Partly, this depends on how the rectification ratio is defined, the base temperature, and the applied temperature gradient across the junction. While some numbers in Table~\ref{table:RR} show very promising values, it must be remembered, that these predictions assume nano-fabrication with high tolerances, and may thus be sensitive to structural changes on the nanometer scale. Furthermore, we here prefer to use a more conservative definition of $RR$ defined as ${|J^+-J^-|}/{{\rm max}(J^+,J^-)}$. This makes $RR$ insensitive to direction and confines it to an interval between zero and unity, where $RR=0$ means no rectifying behavior and $RR=1$ implies completely unidirectional transport. 

\begin{table}[!ht]
\begin{tabular}{m{3.3cm} m{1.7cm} m{2.2cm}} 
\toprule
System & Def. & $RR$ (\%) \\                
\midrule[\heavyrulewidth]
 Y-junction, \mbox{pristine}, Ref.~\mbox{[\!\!\citenum{Yjunction}]} &  ${\displaystyle \frac{J^{+}-J^{-}}{J^{-}}}$  & $\approx38$ \mbox{(SLG)}  $\approx {63}$ \mbox{(BLG)}   \\               
3-Layer/pristine Ref.~\mbox{[\!\!\citenum{Xu_2014}]} & ~~${\displaystyle{J^{-}/J^{+}}}$  & $\approx 40$   \\ 
defected/pristine, Ref.~\mbox{[\!\!\citenum{defect_12}]} & ${\displaystyle\frac{J^{+}-J^{-}}{J^{-}}}$  &$\approx 80$ \\ 
pattern/pristine, Ref.~\mbox{[\!\!\citenum{Series_diode}]} & ~${\displaystyle R^{-}/R^{+}}$ & $\le 120$ \\ 
asymmetric \mbox{ribbons}, Ref.~\mbox{[\!\!\citenum{Hu_2009}]} & ${\displaystyle\frac{K^{+}-K^{-}}{K^{-}}}$ & $\le 120$  \\ 
asymmetric \mbox{ribbons}, Ref.~\mbox{[\!\!\citenum{NuoYang_asym}]} & ${\displaystyle\frac{J^{+}-J^{-}}{J^{-}}}$ & $\le 350$  \\
grain\mbox{ }boundary, Ref.~\mbox{[\!\!\citenum{Cao_grainbound}]} & ${\displaystyle\frac{J^{+}-J^{-}}{J^{-}}}$ & $\le 160$  \\
\bottomrule
\end{tabular}
\caption{Rectification Ratio ($RR$) defined for different geometries of graphene-based thermal rectifiers in literature. Quantities $J$, $K$ and $R$ represent heat flux, thermal conductivity and resistivity, respectively.}
\label{table:RR}
\end{table}

The remainder of this paper is organized as follows. To demonstrate that thermal rectification can be achieved without detailed nano-patterning, we consider a simple thermal junction composed of a short and a long graphene nano ribbon connected via a thermal weak link, and review the basic physical explanation of the origin of this effect in terms of temperature dependent vDOS-overlap. Although we can attribute the effect to the behavior of the overlap of the vDOS, we show that the conventional measure of the overlap is here too simple to quantitatively explain the large observed rectification. Thereafter, we present a simplified system made adjustable using an external force, e.g. electrostatics.

\begin{figure}[!ht]
 \includegraphics[width=\linewidth]{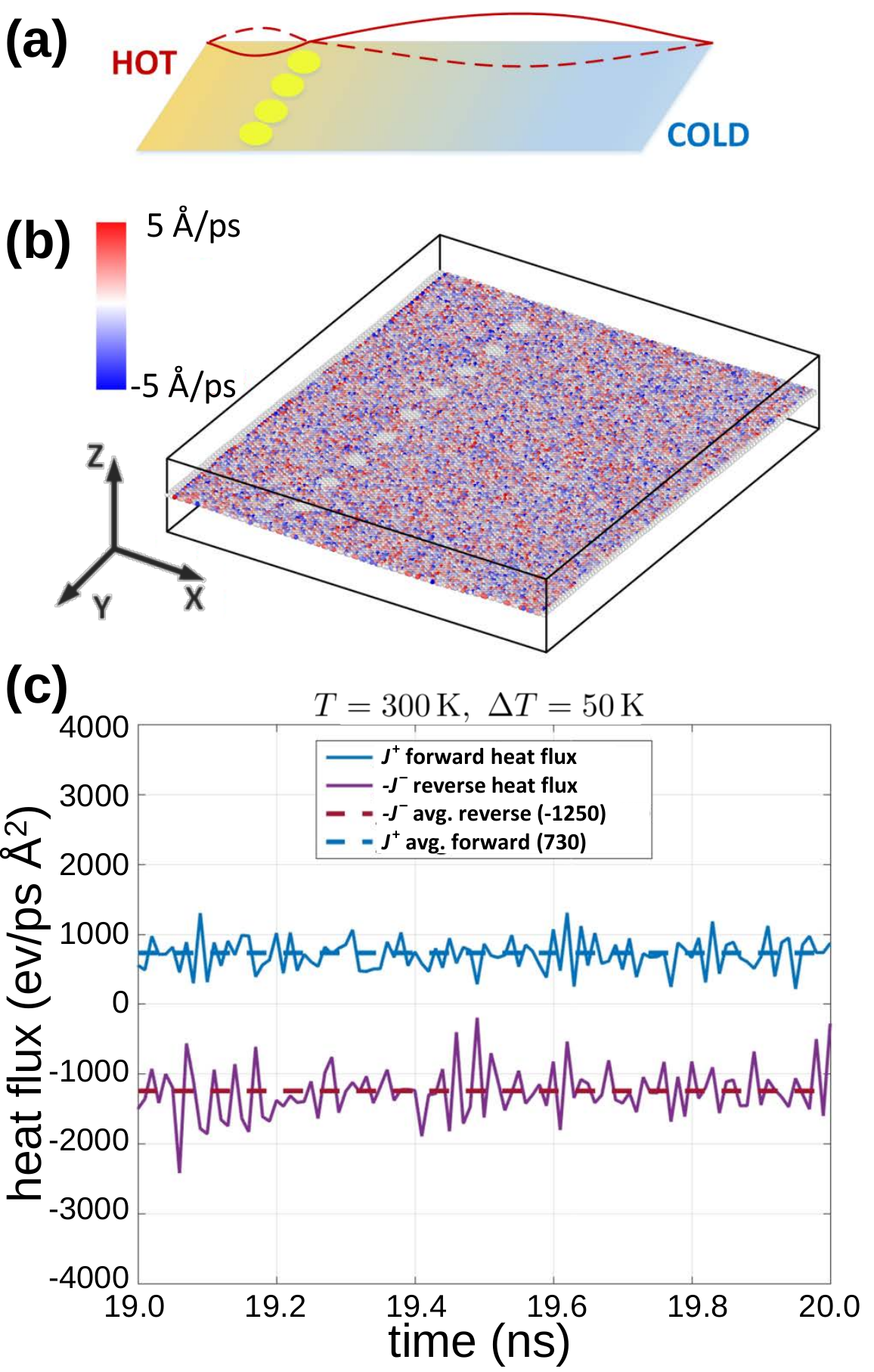}
  \caption{Thermal rectification in an asymmetric graphene junction composed of a short and a long section {\bf (a)}. {\bf (b)} Asymmetric junction made from a $22\,{\rm nm}\times22\,{\rm nm}$ graphene sheet clamped at two ends. A phononic junction is formed by an equidistantly spaced array of pinned circular regions having diameter $d=10$~\AA. The brightly colored areas are clamped sections, i.e. sections with zero velocity and zero force. {\bf (c)} Forward ($J^{+}$) and reverse (-$J^{-}$) heat fluxes when biased with ($T_{\rm L}$, $T_{\rm R}$)=(350~K, 250~K) and (250~K, 350~K), respectively. A clear rectification behavior is observed i.e. $J^{-}>J^{+}$.}
  \label{fig:long-short}
\end{figure}

{\bf Junction thermal rectification}. We consider first a simple junction geometry as shown in Figure~\ref{fig:long-short}(a-b), consisting of a 22~nm$\,\times\,$22~nm graphene sheet clamped at two ends, interrupted by a thermal weak link made from an equidistantly spaced array of pinned circular regions. Using molecular dynamics (MD), the average heat flux flowing through the structure when the two clamped ends are at different temperatures was recorded. Details on the simulations and computational procedures can be found in the Supplementary Information (SI) to this paper.

Prior to imposing pinned sections on the graphene nano ribbon, the thermal conductivity of the graphene sheet is calculated using Non-Equilibrium Molecular Dynamics (NEMD)~\cite{NuoYang_iop2016} for calibration purposes.  Immediately after and before the left and right clamped ends, hot and cold regions are kept at 320~K and 280~K, respectively. Calculating the heatflux $J$, the thermal conductivity $\kappa$ is determined from
\begin{equation}
J_{i} = -\kappa_{ij}\frac{\partial T}{\partial x_{j}}.
\end{equation}
Values of $\kappa=570$~W/m\,K and $\kappa=360$~W/m\,K are obtained using modified Tersoff~\cite{Lindsay_2010} and CH-AIREBO~\cite{AIREBO} potentials, respectively. These values are in close agreement with those obtained using MD for graphene nano ribbons of similar sizes~\cite{Hu_jap2010,Guo2015}.  As the length of the ribbon is smaller than the intrinsic phonon mean free path (MFP) in graphene, $L_{\rm MFP}=600$~nm~\cite{Pop_MRS2012}, a difference between thermal conductivity of a 22~nm$\,\times\,$22~nm in our model and a 20~nm$\,\times\,$20~nm nanoribbon in Refs.\citenum{Hu_jap2010} and\citenum{Guo2015} is expected. The higher thermal conductivity calculated by using the Tersoff potential is attributed to the stiffer C-C bonding in the Tersoff potential as opposed to the AIREBO potential. Stiffer bonds gives a higher group velocity for acoustic phonons and, consequently, a larger thermal conductivity. Finally, it should be pointed out that classical MD neglects quantum corrections, which can be rather large for graphene due to its large Debye temperature~\cite{Quant_Tewary,Quant_Jiang}. However, we are here interested in demonstrating a qualitative phenomenon and derive a ratio between heat-fluxes. Hence, for our purposes, classical MD suffices.

Figure~\ref{fig:long-short}(c) shows the result of simulating the phononic junction of Figure~\ref{fig:long-short}(a-b) where the terminals where kept at temperatures $T_{\rm L}=350$~K, $T_{\rm R}=250$~K and vice versa. A clear rectifying behavior with heat fluxes $J^\pm$ obeying $J^{-}>J^{+}$ and a rectification ratio of $\approx 41$~\% is observed. This rectification ratio corresponds to the 160~\% ratio of reverse to forward heat fluxes, i.e. $J^{-}/J^{+}=1.6$. Changing the pinning diameters to 20~\AA~gives similar results. Results for similar simulations at other base temperatures are shown in Table~\ref{table:22nm}. Interestingly, we find values which are strongly temperature dependent and non-monotonic with increasing temperature. 
\begin{figure}[t]

  \includegraphics[width=\linewidth]{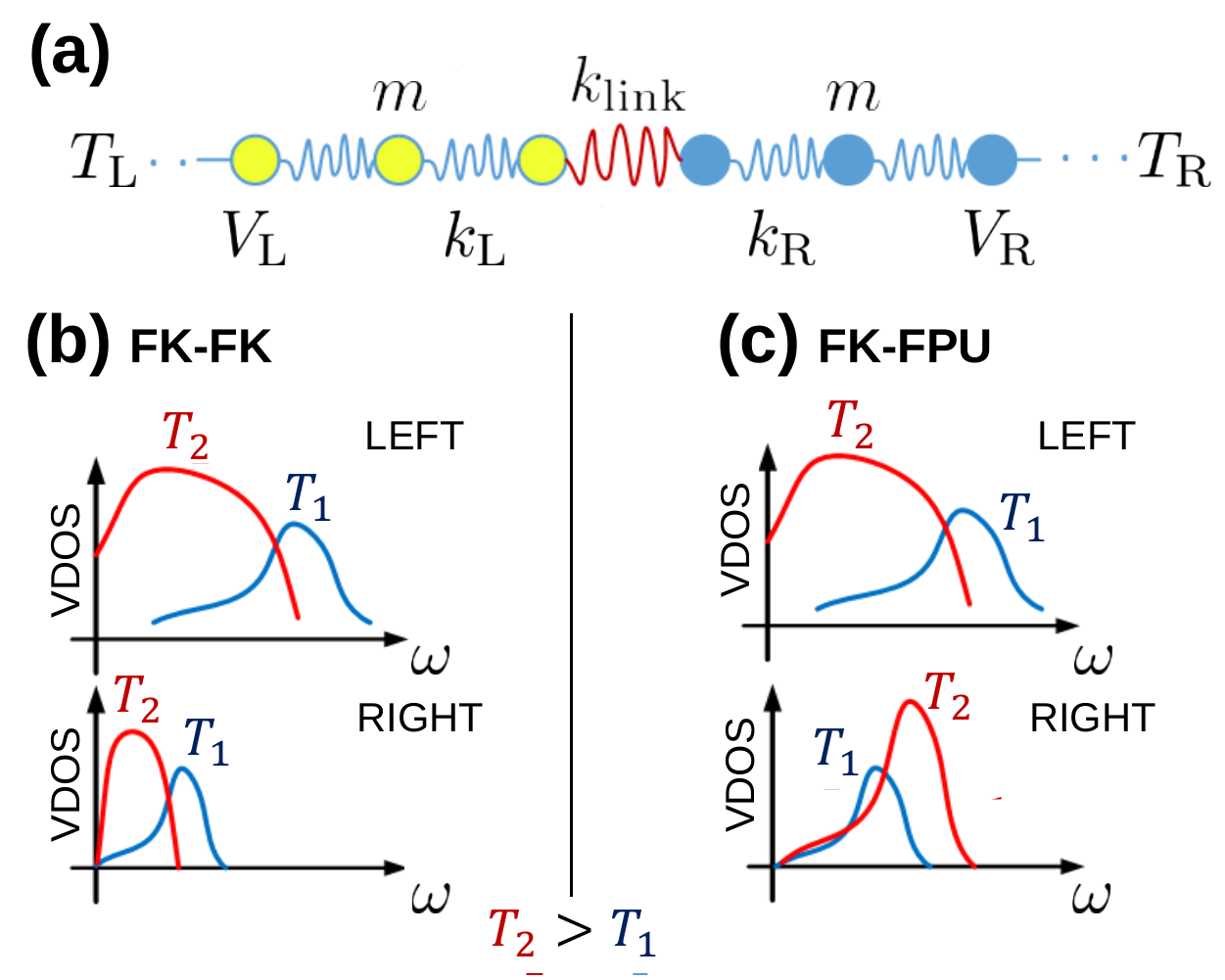}
  \caption{{\bf (a)} One dimensional model of a thermal junction composed of lattices with different potentials joined by a weak link modelled as a spring with stiffness $k_{\rm link}$.  {\bf (b)} When both potentials are of Frenkel-Kontorova (FK) type, stiffer springs and deeper potential for the left section lead to different spectrum for the left section compared to the right. Although for both sections the spectra are red-shifted with increasing temperature, they do so at different rates and in frequency ranges. This leads to asymmetry in the overlap of the vDOS. For example if $T_{\rm L}=T_{2}>T_{\rm R}=T_{1}$, there is a larger overlap hence $J^{+} > J^{-}$. {\bf (c)} When the right hand lattice is modelled by Fermi-Pasta-Ulam (FPU) potential. Since the temperature behavior of the FPU is opposite to that of the FK potential, frequency shifts are in opposite directions with temperature, thereby enhancing further the rectification ratio $RR$. }
  \label{fig:1dmodel}
\end{figure}

\begin{table}[!t]
\begin{tabular}{c c c c} 
\toprule
$T_{\rm L,R}$~(K) & $J^{+}$,$J^{-}$   &  $RR$ (\%) & $S^-/S^+$\\
    & (eV/ps\,\AA$^2$) & $\frac{|J^+-J^-|}{{\rm max}(J^+,J^-)}$  \\
  
\midrule[\heavyrulewidth]
$200\pm50$ & 305, 452  & 33  & 1.0027\\ 
$250\pm50$ & 632, 663  & 4.7  & 1.0004\\ 
$300\pm50$ & 730, 1246  & 41  & 1.0005\\ 
$350\pm50$ & 1365, 1447  & 5.7  & 1.0015\\ 
\bottomrule
\end{tabular}
\caption{Heat fluxes $J^\pm$, rectification ratios $RR$ and vDOS overlap ratios [see Eq.~\eqref{eq:S}] at different thermal biases and base temperatures for the simple junction system in Fig.~\ref{fig:long-short}. }
\label{table:22nm}
\end{table}
To understand the mechanism of rectification in this system, it is instructive to first recall the well studied model of a nonlinear one dimensional chain of oscillators connected by an interface layer (or interface bond)~\cite{Li_2004, Terraneo_2002, Lepri_1997, Nuo_massgrad} which is shown in Figure ~\ref{fig:1dmodel}(a). If the elastic properties on the two sides of the junction differ, by using for instance different spring constants and different strength of nonlinear potential, as a result the corresponding density of vibrational modes will differ. In the absence of inelastic scattering, the vDOS must have an overlap for phonons to cross the junction.

Following the discussion of G. Benenti~\cite{Benenti2016}and B. Li et al.~\cite{Li_2004} we consider a system which is composed of two sections modeled as atoms connected by springs, and a Hamiltonian 
$H=H_{\rm L}+H_{\rm R}$. The left and right sections are weakly linearly coupled with a single link ${k}_{\rm link}$ yielding a total Hamiltonian $H=H_{\rm L}+H_{\rm R}+H_{\rm int}$. The necessary nonlinearity is introduced through adding a Frenkel-Kontorova (FK) co-sinusoidal potential to each section resulting in~\cite{Li_2004}
\begin{eqnarray}
H_{\rm L,R} &=& \sum_{i}\left[{\frac {P_{i}^2}{2m_{i}}}+\frac{1}{2}k_{\rm L,R}(x_{i}-x_{i+1}-a)\right.\nonumber \\ && -\left.\frac{V_{L,R}}{(2\pi)^2}\cos(2\pi x_{i})\right].
\end{eqnarray}
Here $V_{\rm L,R}$ are the depths of the FK potential, $x_i$ and $a$ are atom positions and the lattice constant, respectively. 

Rectifying behavior requires a combination of nonlinearity and broken symmetry. Symmetry breaking is obtained by setting $V_{\rm L}>V_{\rm R}$ and $k_{\rm L}>k_{\rm R}$. This means that the atoms in the left section are connected by stiffer springs and are exposed to a deeper FK potential as opposed to the right section. The FK-potential is a softening potential, i.e. high amplitude vibrations have lower frequencies than low amplitude ones. Hence, at low temperatures they oscillate with higher frequencies than at high temperatures. Thus with increased temperatures the vDOS moves toward lower frequencies [see Figure~\ref{fig:1dmodel}(b)]. The same trend exists for the right section of the junction however with different rate as well as different frequency. As springs on the right are softer than those on the left, the vDOS on the left is predominantly covering lower frequencies. This explains why for $T_{R}>T_{L}$ the vDOS overlap is smaller than when $T_{\rm R}<T_{\rm L}$ [see Figure~\ref{fig:1dmodel}(b)]. 

\begin{figure*}[!t]
\includegraphics[width=\linewidth]{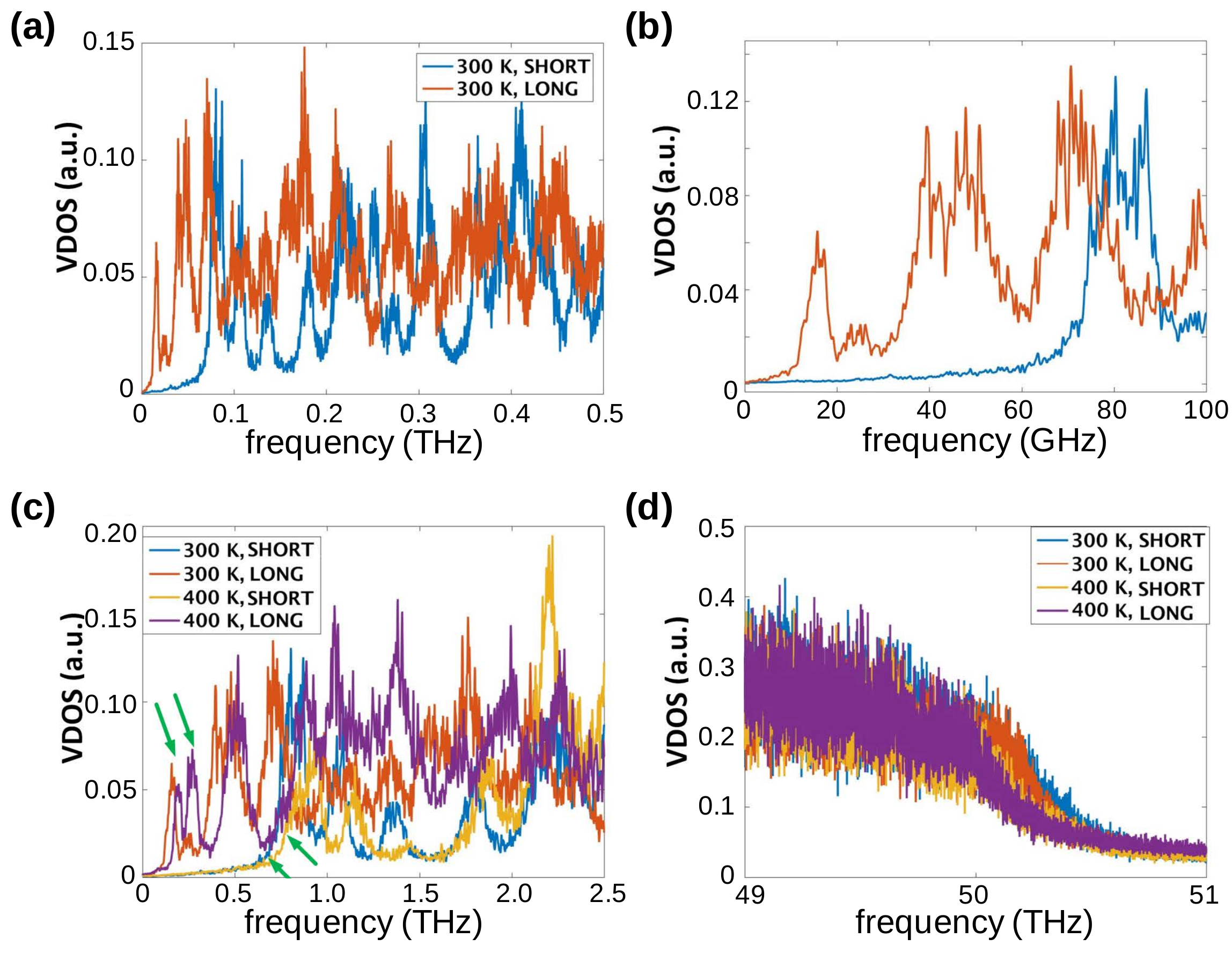}
 
  \caption{Low frequency density of vibrational states for temperatures of 300 K and 400 K. {\bf(a)} Spectra for long and short sections of device at 300 K, {\bf(b)} Magnified section of the spectra shown at {\bf(a)} for 0-100 GHz frequency range. {\bf(c)} Spectra for both sections at 300 K and 400 K revealing stiffening of the sections by increasing the temperature which is manifested as frequency up shift (marked by arrows), and {\bf(d)} the high frequency section of the spectra from 49 THz - 51 THz shows softening of the optical phonon modes, hence a frequency down shift is observed for both sections by increasing the temperature.}
  \label{fig:vdos}
\end{figure*}

To enhance the rectification, it is favorable to use hardening and softening potentials on either sides. An example is to replace the FK-potential in the right with a Fermi-Pasta-Ulam (FPU) type system. The nonlinearity in the FPU-$\beta$ model is an extra anharmonic term like $\beta x^4$ in the Hamiltonian with $\beta>0$. Due to this term, the behavior of the vDOS with temperature is opposite on the right and left sides. Hence, for an FPU-$\beta$ potential~\cite{BLi_2dmodel} the increase of temperature shifts the spectrum to higher frequencies according to $\omega \propto (\beta T)^{1/4}$. This means that vDOS of the left and right sections of the thermal junction move in opposite directions with increasing temperatures and, consequently, to enhanced thermal rectification [see Figure~\ref{fig:1dmodel}(c)]. 

We now return to the 2D-system of Figure~\ref{fig:long-short}. We consider the area with pinned circles as the weak link between the left and right sections creating the symmetry breaking. To test if the rectification mechanism is similar to the 1D-systems we have studied the vDOS $D_{\rm L,R}(\omega)$ corresponding to the left and right sections of the nanoribbon at different temperatures (Figure~\ref{fig:vdos}). 

First we inspect the low-frequency part of $D(\omega)$, shown for the two sections at $T=300$~K in panels (a) and (b).  As anticipated, and in contrast to the case of spring-mass models where the spring constant determines the bottom of the band, here the lateral confinement will determine the band bottom. Indeed, regardless of temperature, the left section (short) is gapped from 0~GHz to 58~GHz, while the right (long) section supports pronounced low frequency modes from 10~GHz and upwards. Hence, we can identify the longer (right) side the soft-spring part and the short side (left) with the stiff-spring part. 

Figure~\ref{fig:vdos}(c) shows how the acoustic parts of the spectra shifts with changing temperature for both sides. For these low frequency portions of the spectra, increasing the temperature causes blue-shift to higher frequencies, which resembles the way in which FPU-$\beta$ model behaves~\cite{BLi_2dmodel}. The nonlinearities responsible for this comes partly from non-linearity of the Tersoff many-body potential~\cite{Lindsay_2010} itself. Partly, it can be attributed to the negative thermal contraction of graphene which stiffens the acoustic ZA-modes. This leads to a situation akin to the FK-FK situation in Fig.~\ref{fig:1dmodel}(b) where both spectra shift in the same direction but by unequal amounts. Qualitatively we indeed find agreement that $J^{-} > J^{+}$ for all investigated cases.

Although giving little contribution to the thermal transport, the high-frequency part of $D_{\rm L,R}(\omega)$, corresponding to the optical phonon branches, is shown in Fig.~\ref{fig:vdos}(d). Here the short section supports high frequency modes up to 50.7~THz even at 50~K. However, by increasing the temperature from 300~K to 400~K we observe a gradual shift to lower frequencies e.g. 50~THz. The same trend is true for the long (right) section of the ribbon although the modes have generally lower frequency spanning from 50~THz - 50.4~THz. This is consistent  with the observed red shift of the G-peak in Raman measurements, and suggests that increased temperature causes softening of the optical modes and reduction of group velocity due to tensile strain build-up~\cite{Hong_2016, strain_iop}. 

As a measure of the degree of overlap, Li et al.~\cite{BLi_PRL2005} defined a phenomenological quantity $S$ as the vDOS overlap
\begin{equation}
S= \frac{\int D_{\rm L}(\omega)D_{\rm R}(\omega)d\omega}{\int D_{\rm L}(\omega)d\omega\int D_{\rm R}(\omega)d\omega},\label{eq:S}
\end{equation}
which they relate to the rectification ratio by comparing $J^-/J^+$ to $S^-/S^+$. To quantitatively support the above mentioned arguments we have calculated the ratio $S^-/S^+$ according to Eq.~\eqref{eq:S} for the simulations presented in Table~\ref{table:22nm}. We obtain $D_{\rm L,R}$ from using the velocity auto correlation function (see the SI). We assume that the left and right section are disconnected by fixing the ribbon which connects these two regions. The ribbon already covers the pinned circles and it will be fixed by zero force and zero velocity. 
Qualitatively we find, in agreement with the simulated heat fluxes, that the overlap is always larger when $T_{\rm R}>T_{\rm L}$. This is reflected in Table~\ref{table:22nm} as $S^-/S^+>1$. However, we find no other quantitative correlations between $S^-/S^+$ and $RR$. This is hardly surprising, considering the inclusion and thermal shift of optical modes in the calculation as well as the presence of localized modes in the spectra which may contribute to $S$ but not to $J$. Hence, we conclude that the overlap $S$ alone is not sufficient to quantify the rectifying behavior. 

\begin{figure}[!t]
 \includegraphics[width=\linewidth]{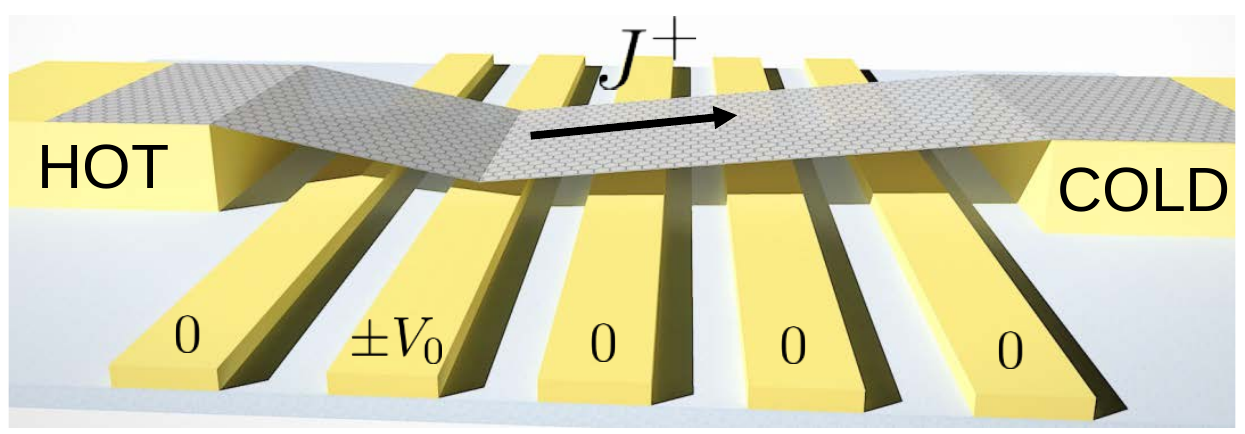}
  \caption{ Proposal for a tunable thermal diode where the sign and value of rectification ratio is controlled by applying electric field to the chosen gate electrode underneath a thermally biased graphene nano ribbon.}
  \label{fig:proposal}
\end{figure}

{\bf Adjustable Heat Rectification}.
Based on the above method of dividing a nano ribbon into a short and long section, we here propose a simpler design with two distinct advantages to other proposals. Instead of partitioning the ribbon in two sections by pinning, we consider applying a force to the nano ribbon in an asymmetric fashion. The applied force could emanate from sources like mechanical actuation (e.g. piezoelectricity, Atomic Force Microscopy (AFM) tip~\cite{OdedHod}, Scanning Tunneling Microscopy (STM) tip~\cite{Neek-Amal2014}) or electrical force due to an electric field from a local back-gate. The first advantage is that no explicit nanopatterning is needed. The second, and more important advantage, is that changing the position and magnitude of the applied force allow for {\it in situ} reconfiguration and control of the rectification. 

\begin{figure}[t]
\includegraphics[width=\linewidth]{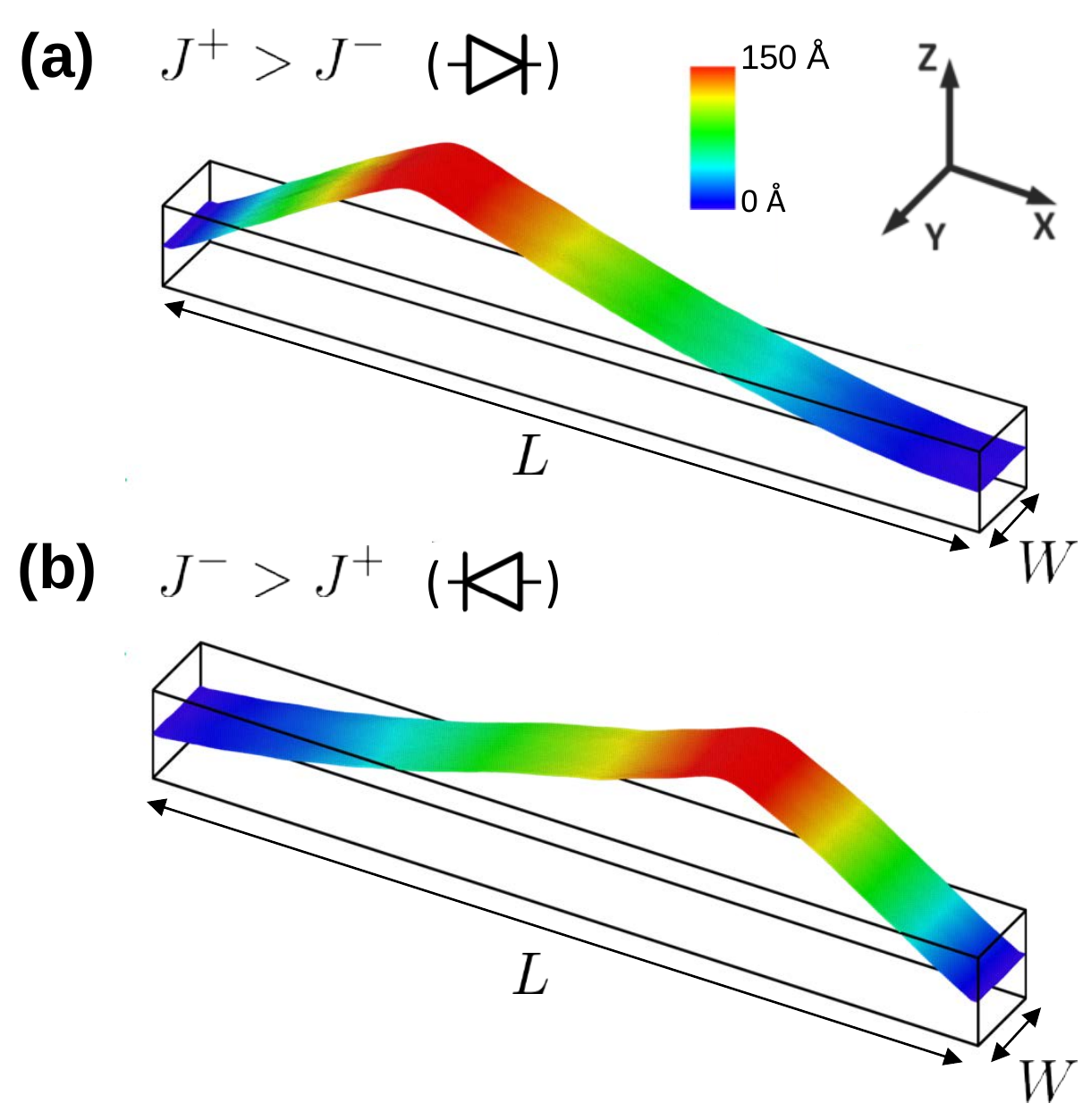}
\caption{Adjustable heat rectification by asymmetric application of electric field-induced force to a graphene nanoribbon of dimensions $L=100$~nm and $W=10$~nm. {\bf(a)} Schematic shows the rectification direction of the heat diode when the force is applied to the left section. {\bf(b)} When the force is applied to the right section, rectification direction is reversed. With the proposed setup in Fig.~\ref{fig:proposal} it is thus possible to realize an \emph{in situ} re-configurable thermal rectifier.}
\label{fig:feld}
\end{figure}

Specifically we envision a setup here as shown in Fig.~\ref{fig:proposal} which is similar to the experimental setup already used to modulate the electron transport properties in carbon nanotubes~\cite{Benyamini_2013}. Applying electric field to the selected electrode can divide the nano ribbon into a short and long section. Here a proof of concept molecular dynamics simulation is performed to demonstrate the rectification as a result of force-induced asymmetry.

 Figure~\ref{fig:feld}(a), shows a simulation set-up of a $100\,{\rm nm}\times 10\,{\rm nm}$ nano ribbon, in which a force is locally applied along the $z$-direction to a narrow strip causing deflection. The asymmetric application of the force divides the strip into a long and short section with unequal tensile stresses. The  magnitude of forces used in the simulations are of the order nN, characteristic for nanoelectromechanical devices.

 The recorded values of average heat fluxes are shown in Table~\ref{table:Field} and they suggest rectification ratios of 35~\% and 42~\% for temperature/gradient values of  $300\,{\rm K}\pm 100\,{\rm K}$ and $300\,{\rm K}\pm50\,{\rm K}$, respectively. If the position of force is changed and it is moved to the right side Figure~\ref{fig:feld}(b), the same values as above are obtained for the rectification ratios however, with the reversed sign. As for the junction rectifier, we also changed the base temperature to 200~K and 400~K respectively. As can be seen from Table~\ref{table:Field}, the value of the rectification ratio changes. In case of $T_{\rm L,R}=400\,{\rm K}\pm 50\,{\rm K}$ and $T_{\rm L,R}=200\,{\rm K}\pm 50\,{\rm K}$, the $RR$ values are 60~\% and 22~\%, respectively. 

\begin{table}[!t]
 \begin{tabular}{c c c c} 
 \toprule
$T_{\rm L,R}$~(K) & $J^{+}$,$J^{-}$   &  $RR$ (\%) & Rect.\\
    & (eV/ps$\,$\AA$^2$) & $\frac{|J^+-J^-|}{{\rm max}(J^+,J^-)}$  & dir.\\
\midrule[\heavyrulewidth]
 $300\pm 50$ & 5191, 2998  & 42 & S$\rightarrow$L\\ 
 $300\pm 100$ & 6451, 4201 & 35 & S$\rightarrow$L\\ 
\midrule
 $200\pm 50$ & 1542, 3903  & 60 & L$\leftarrow$S\\ 
 $400\pm 50$ & 3822, 4931  & 22 & L$\leftarrow$S\\ 
 \bottomrule
\end{tabular}
\caption{Heat fluxes $J^\pm$, rectification ratios $RR$ and rectification direction (Short$\rightarrow$ Long) for the tunable configuration in Fig.~\ref{fig:feld}. The top two rows correspond to configuration in panel (a), while the bottom two correspond to panel (b).}
\label{table:Field}
\end{table}

To verify that the rectification is indeed induced from asymmetric length of the nano ribbon section, we also applied the force symmetrically, dividing the ribbon into two equal partitions. The resulting flux differences for imposed temperature differences of $\Delta T=50\,{\rm K}$ and $\Delta T=100\,{\rm K}$ corresponded to $RR=6$~\% and $RR=1$~\%, respectively. This corroborates that the large asymmetry in the values of heat flux stems from the strain induced asymmetry in the nano ribbon.

The significance of this work lies in its flexibility of changing the value and sign of heat rectification without resorting to a newly nano-patterned and/or functionalized graphene. It is also possible to generalize the proposed principle to other two dimensional materials using not only the electric field but also magnetic field with proper electrode designs~\cite{DS_AI_notpublished}. 

{\bf Conclusions}.
Based on molecular dynamics studies of non equilibrium heat transport in graphene nano ribbons we showed that the thermal rectification ratio can be simply adjusted both in value and sign using a force to divide the graphene into strained long and short sections. Starting from a simple structure in which long and short sections are weakly linked by pinned areas, we showed how this partition lead to the necessary temperature dependent asymmetric vibrational density of states required for thermal rectification. 

The importance of this observation is reflected in the simplicity in implementing an {\it in situ} tunable heat diode, as examplified by using an array of local back gates which can selectively induce asymmetric stress in a suspended graphene membrane. Our findings show that the values of heat flux and the rectification ratio can be controlled on-the-fly while the device is operating. Consequently, to achieve rectification it is not necessary to fabricate elaborate asymmetric shapes, {\it i.e.} patterning arrays of holes, defects, inducing grain boundaries or the requirement of making a hetero-structure~\cite{Hong_2016} of pristine graphene with other kinds of graphene (multi-layer~\cite{Yjunction,Xu_2014}, defected~\cite{defect_12},patterned~\cite{NuoYang_asym,Hu_nl09,Series_diode,Yang_2009,YanWang_nl14}). Thus, the combination of simplicity of implementation, and tunability of rectification ratio and sign, opens up for more controlled experiments and developments in the field of 2D-phononics. 

{\bf Acknowledgments.} We thank Prof. Paul Erhart and Prof. Nuo Yang for helpful suggestions and communications. Access to supercomputing facilities of The Swedish National Infrastructure for Computing (SNIC) and C3SE is greatly acknowledged. We acknowledge financial support from the Swedish Research Council (VR). 

\bibliography{Thermobib}
\end{document}